\documentclass[aps,prd,12pt,nofootinbib]{revtex4}

\usepackage{amsfonts}
\usepackage{amsmath,epsfig,bm}
\usepackage{graphicx}
\usepackage{color}

\usepackage[setpagesize=false]{hyperref}

\usepackage{tabularx}  %for table formatting 
\newcolumntype{s}{>{\centering\arraybackslash\hsize=.666\hsize}X} 
\newcolumntype{b}{>{\centering\arraybackslash\hsize=1.333\hsize}X} 

\usepackage{epsfig}

\newcommand{\be}{\begin{equation}}
\newcommand{\ee}{\end{equation}}
\newcommand{\ba}{\begin{eqnarray}}
\newcommand{\ea}{\end{eqnarray}}
\newcommand{\bd}{\begin{displaymath}}
\newcommand{\ed}{\end{displaymath}}

\def\tthalf{{\textstyle{\frac{3}{2}}}}

\def\Zpt{Z_{\rm pt}}
\def\Zex{Z_{\rm ex}}
\def\Ppt{P_{\rm pt}}
\def\Pex{P_{\rm ex}}
\def\spt{s_{\rm pt}}
\def\sex{s_{\rm ex}}
\def\npt{n_{\rm pt}}
\def\nex{n_{\rm ex}}
\def\ept{\epsilon_{\rm pt}}
\def\eex{\epsilon_{\rm ex}}

\begin{document}

\title{Matching Excluded Volume Hadron Resonance Gas Models and Perturbative QCD to Lattice Calculations}

\author{M. Albright, J. Kapusta and C. Young}
\affiliation{School of Physics \& Astronomy, University of Minnesota, Minneapolis, MN 55455,USA}
\
\date{April 29, 2014}

\begin{abstract}
We match three hadronic equations of state at low energy densities to a perturbatively computed equation of state of quarks and gluons at high energy densities.  One of them includes all known hadrons treated as point particles, which approximates attractive interactions among hadrons.  The other two include, in addition, repulsive interactions in the form of excluded volumes occupied by the hadrons.  A switching function is employed to make the crossover transition from one phase to another without introducing a thermodynamic phase transition.  A chi-square fit to accurate lattice calculations with temperature $100 < T < 1000$ MeV determines the parameters.  These parameters quantify the behavior of the QCD running gauge coupling and the hard core radius of protons and neutrons, which turns out to be $0.62 \pm 0.04$ fm.  The most physically reasonable models include the excluded volume effect.  Not only do they include the effects of attractive and repulsive interactions among hadrons, but they also achieve better agreement with lattice QCD calculations of the equation of state.  The equations of state constructed in this paper do not result in a phase transition, at least not for the temperatures and baryon chemical potentials investigated.  It remains to be seen how well these equations of state will represent experimental data on high energy heavy ion collisions when implemented in hydrodynamic simulations.
\end{abstract}

\maketitle

\parindent=20pt

\newpage

\section{Introduction}
\label{Intro}

The equation of state of Quantum Chromodynamics (QCD) at finite temperature is studied theoretically in a variety of ways.  Starting from {\it low} temperatures one has a dilute gas of pions and nucleons.  With increasing temperature hadron resonances are created and contribute to the equation of state.  If the spectrum of resonances increases exponentially with mass then one reaches a Hagedorn limiting temperature which experiments and models suggest is about 160 MeV.  This conclusion is based on the treatment of hadrons as point particles, which they are not.  Starting from extremely {\it high} temperatures one can use perturbation theory to calculate the equation of state because QCD has the property of asymptotic freedom whereby the effective gauge coupling decreases logarithmically with temperature.  As the temperature is lowered the coupling eventually becomes large and perturbation theory is no longer useful.  The only reliable approach for all temperatures is to do numerical calculations with lattice QCD. 

The goal of this paper is to find a means for switching from a hadron resonance gas at low temperature, preferably treating the hadrons not as point particles but as extended objects, to a plasma of weakly interacting quarks and gluons at high temperature.  We will construct a switching function that does just that.  The parameters will be adjusted to fit the lattice equation of state at zero chemical potentials.  Then the model can make parameter-free predictions for both finite temperature and chemical potentials.  Lattice calculations at finite chemical potentials face well-known problems, but comparison to one of them at a baryon chemical potential of 400 MeV is quite good.   The equation of state constructed in this paper can be used in hydrodynamical models of high energy heavy ion collisions.  It has the advantage that at the moment of freeze-out from fluid behavior to individual hadrons, one will know the chemical abundance of all the hadrons which then can either be compared to experimentally observed abundances or used as the initial condition for a cascade after-burner.

The outline of this paper is as follows.  In section \ref{point} we will compare the hadron resonance model of point particles and the most recent calculations of perturbative QCD to the lattice equation of state to illustrate the problem we are addressing.  In section \ref{excluded} we will review and extend two versions of the excluded volume model which take into account the extended spatial size of hadrons.  In section \ref{sec:switch} we will construct a switching function, and adjust its parameters and the other parameters in the model by doing a chi-square fit to both the pressure and the trace anomaly/interaction measure.  The resulting parameters provide physical information, such as the size of hadrons and one optimum way to choose the scale of the running gauge coupling as a function of temperature and baryon chemical potential.  In section \ref{chemical} we will compare with lattice results at a baryon chemical potential of 400 MeV.  Our conclusions are  contained in section \ref{conclusion}.

\section{Hadron Resonance Gas and Perturbative QCD}
\label{point}

The equation of state of the hadronic phase is usually assumed to be a hadron resonance gas where all observed, and sometimes extrapolated, hadrons are included as free non-interacting point particles.   According to the arguments by Dashen, Ma and Bernstein \cite{DMB1}, this is a reasonable way to include attractive interactions.  (Repulsive interactions will be addressed in the next section.)  Each hadronic species labeled by $\alpha$ contributes to the pressure as
\be
P_{\alpha} =  (2s_{\alpha}+1) \int \frac{d^3p}{(2\pi)^3} \frac{1}{{\rm e}^{\beta (E_{\alpha}(p)-\mu_{\alpha})} \pm 1}
\ee
The sign is chosen according to whether the hadron is a boson or a fermion.  The inverse temperature is denoted by $\beta$ and $\mu_{\alpha}$ is the chemical potential.  Formulas for the energy density, entropy density, and for the conserved quantum numbers are standard.  Following a well-trodden path, we include all hadrons appearing in the most recent Particle Data Group compilation.  For completeness, and for use by others, we provide a table in Appendix A.

The equation of state of the quark-gluon plasma is calculated using perturbation theory in the gauge coupling.  Many papers have contributed to this endeavor since the first papers in the late 1970's.  Here we use the latest results which include terms up to order $\alpha_s^3 \ln \alpha_s$.  The formula for the pressure is given in Appendix B.  We assume 3 flavors of massless quarks.  There are two issues with obtaining accurate numerical results.  First, it was observed early on that the series in $\alpha_s$ is oscillatory, so that at non-asymptotic temperatures the results depend to some degree on where the series is terminated.  Second, one has some freedom in choosing the renormalization scale $M$ for $\alpha_s$.  In ref. \cite{Kapusta1979} it was suggested to choose $M^2$ roughly equal to the average three-momentum of the quarks and gluons.  For massless particles with quark chemical potential $\mu_q = \mu/3 = 0$, one finds $M \approx 3T$, the exact coefficient depending on whether they are bosons or fermions.  For massless particles with $T=0$, one finds $M \approx \mu_q$.  Another commonly used argument for the choice of scale is that $M = \pi T$ since that is the smallest Matsubara frequency.  We shall choose
\be
M = C_M \sqrt{(\pi T)^2 + \mu_q^2}
\label{scaleM}
\ee
and adjust the coefficient $C_M$ to best represent the lattice results.  What is important here is the relative proportion of $T$ and $\mu_q$, which is chosen on the basis of the above arguments.  The quantity labeled by $t$ which enters into the solution of the 3-loop beta function for the running coupling would usually be taken to be 
$t=\ln(M^2/\Lambda_{{\overline{MS}}}^2)$.  This results in a divergence of the running $\alpha_s$ at small but finite values of the temperature and chemical potential, the famous Landau pole.  In reality one would expect $\alpha_s$ to remain finite even at zero energy scale, although its value is most likely gauge-dependent.  To address this problem we choose
\be
t = \ln \left(C_S^2 + M^2/\Lambda^2_{\overline{MS}} \right)
\ee
where $C_S$ is a constant used to soften the Landau pole; along with $C_M$ it will be adjusted to represent best the lattice results.

The lattice results at zero chemical potential to which we compare were reported in \cite{Borsanyi2010}.  They included 2+1 flavors of quarks (strange quark heavier than up and down quarks).  The temperature range sampled was from 100 to 1000 MeV, extending even beyond the highest temperatures expected at CERN's Large Hadron Collider (LHC).  Figure 1 shows the pressure divided by $T^4$.  The hadron resonance gas represents the lattice result very well up to about $T = 200$ MeV and then greatly exceeds them.  If we had included a full exponential spectrum of hadronic states, with level density proportional to $\exp(m/T_H)$ where $T_H = 160$ MeV is the Hagedorn temperature, the pressure would either end at a finite value or diverge at $T_H$, depending on the pre-exponential factor.  This doesn't happen here because we include a very large but still finite number of hadronic states.  The perturbative QCD result represents the lattice result very well down to a temperature of about 200 MeV.  It appears from this figure that doing a little matching between the two limiting forms of the pressure in the vicinity of 200 MeV would achieve our goal.

\begin{figure}
\begin{center}
\includegraphics[width=0.8\linewidth]{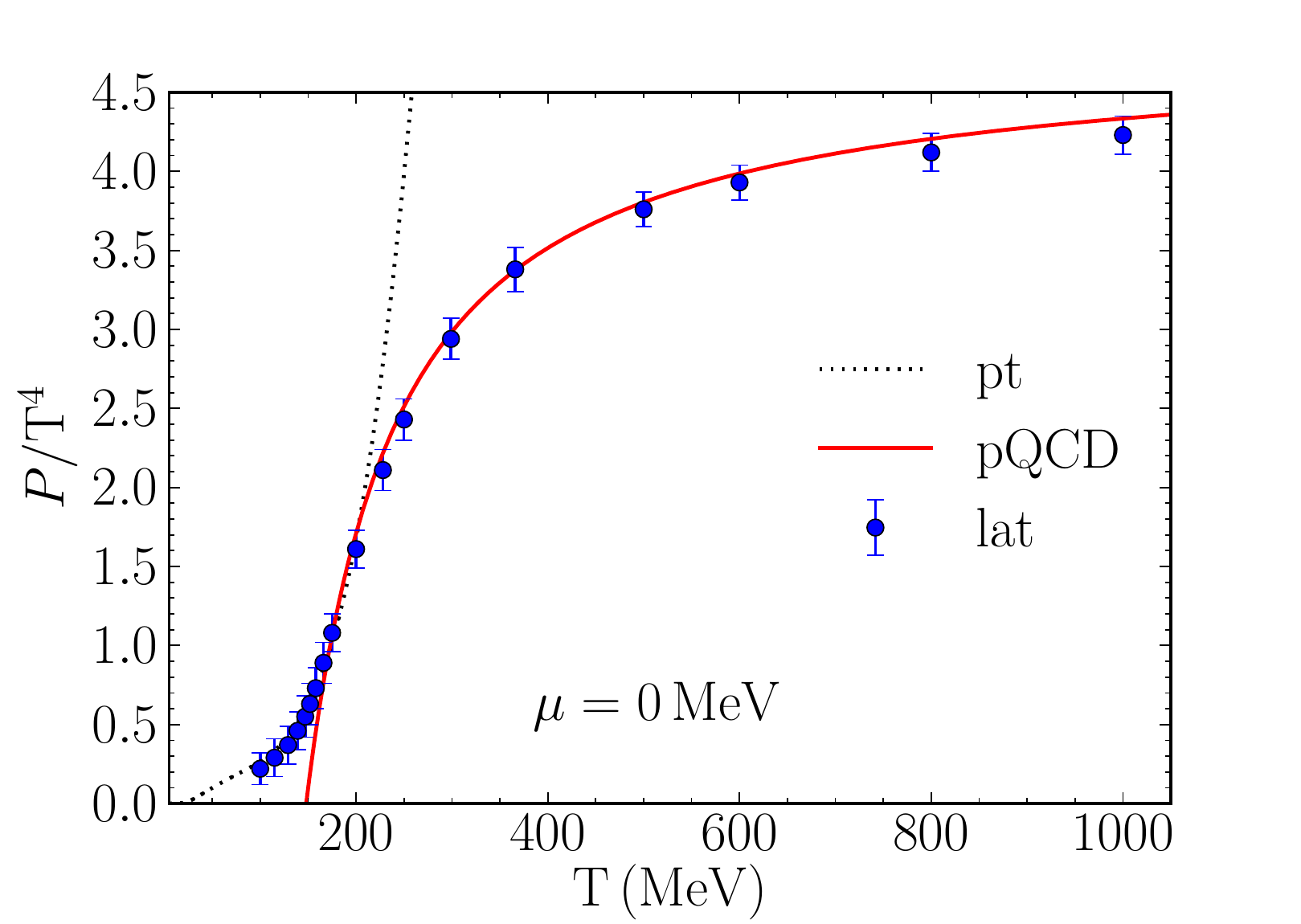}
\caption{Pressure normalized by $T^4$.  The dotted curve represents the parameter-free, point particle hadron resonance gas.  The solid curve represents perturbative QCD with 2 parameters adjusted to fit the lattice result taken from \cite{Borsanyi2010}.}
\end{center}
\label{fig:pressureHQ}
\end{figure}

\begin{figure}
\begin{center}
\includegraphics[width=0.8\linewidth]{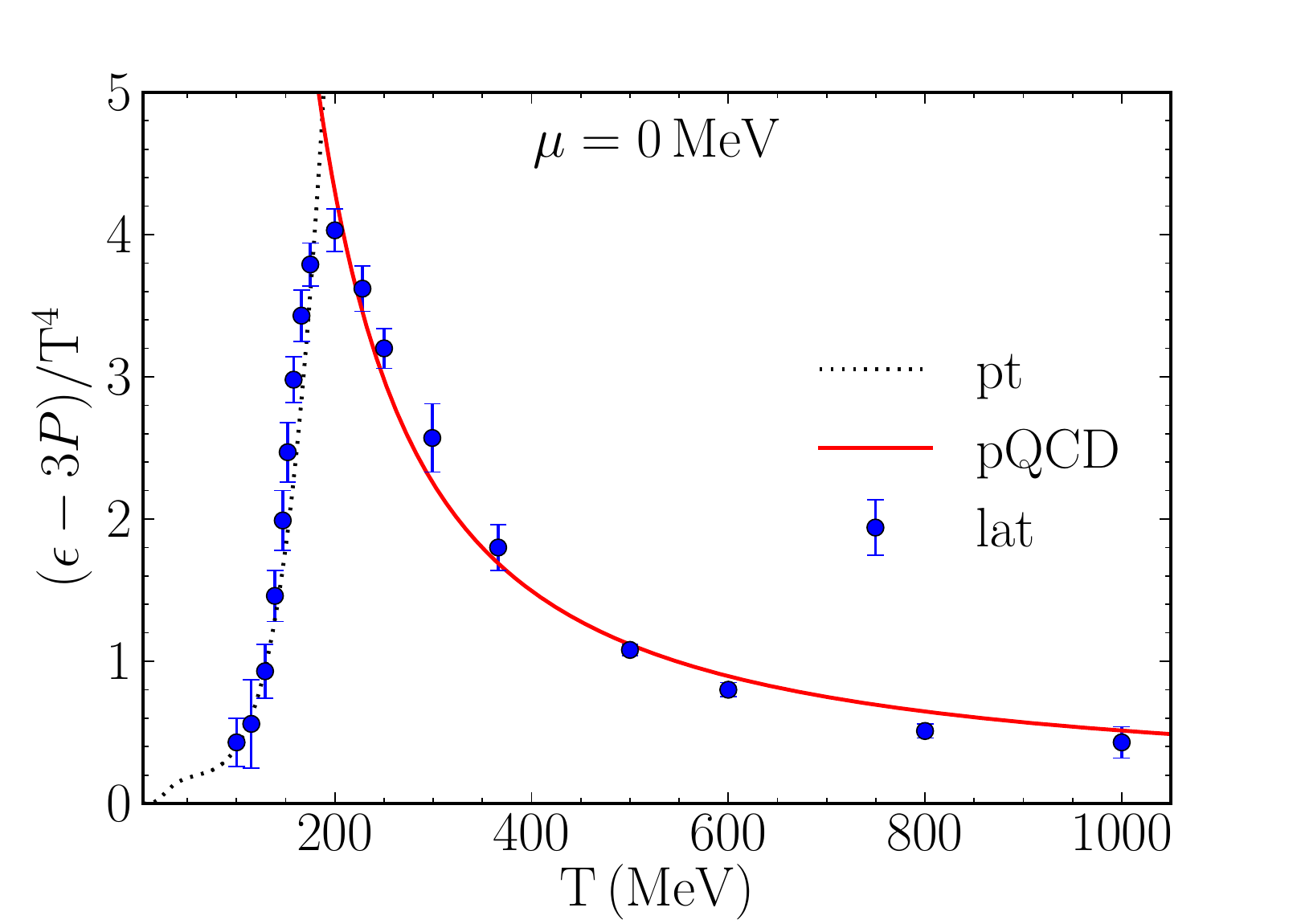}
\caption{Trace anomaly normalized by $T^4$.  The dotted curve represents the parameter-free, point particle hadron resonance gas.  The solid curve represents perturbative QCD with 2 parameters adjusted to fit the lattice result taken from \cite{Borsanyi2010}.}
\end{center}
\label{anomalyHQ}
\end{figure}

Figure 2 shows the trace anomaly, sometimes also called the interaction measure, of $(\epsilon - 3P)/T^4$.  The hadron resonance gas represents the lattice result very well up to a temperature of about 150 MeV and then greatly exceeds it.  This is due to an increasing number of massive hadronic states with increasing temperature, massively diverging from a free massless gas which has $\epsilon = 3P$.  The perturbative QCD result represents the lattice result very well down to a temperature of about 220 MeV.  It also massively deviates at lower temperature because the renormalization group running coupling is becoming large, reflecting the intrinsic QCD scale $\Lambda_{\overline{MS}}$.  Between these two limiting contributions there is a cusp around 190 MeV.

The perturbative QCD parameters were chosen by doing a chi-square combined fit to the pressure and the trace anomaly for $T > 200$ MeV.  For definiteness we fixed $\Lambda_{\overline{MS}} = 290$ MeV, but it should be noted that the choice is irrelevant since the value of $C_M$ can be adjusted accordingly.  The result of the fit is $C_M = 3.293$ and $C_S = 1.509$ with a chi-squared per degree of freedom of 1.397.

It would appear from these results that one could just terminate the hadron resonance gas contribution somewhat below 200 MeV and the perturbative QCD contribution somewhat above, and find an interpolating function to fill in the middle.  The problem is that eventually one will find that some $n$'th order derivative of $P$ with respect to $T$ will become discontinuous at each of the matching points, leading to a phase transition of order $n$.  This is unacceptable.  We tried various matching functions, arguing that if $n$ is large enough it would have no practical effects for use in modeling heavy ion collisions, but we did not succeed.  In addition, one would have to do this interpolation as a function of $\mu$, and the equation of state for arbitrary $T$ and $\mu$ is not known from lattice calculations.  

\section{Excluded Volume Models}
\label{excluded}

Hadrons are not point particles, and repulsive interactions can be implemented via an excluded volume approximation whereby the volume available for the hadrons to move in is reduced by the volume they occupy, as first suggested in \cite{HR1,H1,GPZ1}.  There are at least two thermodynamically self-consistent versions of this model.  Here we extend one of these models, referred to as model I \cite{KO1}, which was originally formulated at finite temperature, to include finite chemical potentials.  Then we compare and contrast it to what we refer to as model II \cite{RGSG1}, which appeared a decade later.  Model II has been compared to experimental data a number of times, such as in \cite{CGSS1} and \cite{YGGY1}.  Our arguments are phrased in terms of classical statistics, or Boltzmann distributions, for clarity of presentation.  However, this is not a limitation, and the extension to quantum statistics is deferred to later.

\subsection{Model I}

In the independent-particle approximation the partition function for a hadron of species $\alpha$ is $V z_{\alpha}$ where $V$ is the total volume of the system and
\be
z_{\alpha} = (2s_{\alpha}+1) \int \frac{d^3p}{(2\pi)^3} {\rm e}^{-\beta (E_{\alpha}(p)-\mu_{\alpha})}
\ee
In the canonical ensemble the total number of particles is fixed, whereas in the grand canonical ensemble only the average is.  Let $n$ denote the total number of species.  The partition function in the grand canonical ensemble in the point particle approximation is
\be
\Zpt = \sum_{N=0}^{\infty} \sum_{N_1=0}^{\infty} \cdot\cdot\cdot \sum_{N_n=0}^{\infty} \delta_{N_1 + \cdot\cdot\cdot N_n,N}
\frac{(V z_1)^{N_1}}{N_1!} \cdot\cdot\cdot \frac{(V z_n)^{N_n}}{N_n!}
\ee
where $N$ is the total number of particles irrespective of species.  The excluded volume approximation being applied here reduces the total volume $V$ by the amount occupied by the $N$ hadrons
\be
V_{\rm ex} = \frac{1}{\epsilon_0} \left[ \sum_{j=1}^{N_1} E_1(p_j) + \cdot\cdot\cdot \sum_{j=1}^{N_n} E_n(p_j) \right]
\ee
The assumption is that the volume excluded by a hadron is proportional to its energy with the constant of proportionality $\epsilon_0$ (dimension of energy per unit volume) being the same for all species.  It is also assumed that hadrons are deformable so that there is no limitation by a packing factor as there would be for rigid spheres, for example.  This is model I.

In the pressure ensemble \cite{HR1,H1} the partition function is the Laplace transform of the grand canonical partition function in volume space.
\be
{\tilde Z}(T,\mu,\xi) = \int dV Z(T,\mu,V) {\rm e}^{-\xi V}
\ee
In the present context the relevant integral is
\be
\int_{V_{\rm ex}}^{\infty} dV (V-V_{\rm ex})^N {\rm e}^{-\xi V} = \frac{N!}{\xi^{N+1}} {\rm e}^{-\xi V_{\rm ex}}
\ee
Then
\be
{\tilde \Zex}(T,\mu,\xi) =  \frac{1}{\xi} \sum_{N=0}^{\infty} \sum_{N_1=0}^{\infty} \cdot\cdot\cdot \sum_{N_n=0}^{\infty} \delta_{N_1 + \cdot\cdot\cdot N_n,N}
\frac{N!}{N_1! \cdot\cdot\cdot N_n!} \left(\frac{{\tilde z}_1}{\xi}\right)^{N_1} \cdot\cdot\cdot \left(\frac{{\tilde z}_n}{\xi}\right)^{N_n}
\ee
where
\be
{\tilde z}_{\alpha} = (2s_{\alpha}+1) \int \frac{d^3p}{(2\pi)^3} {\rm e}^{-(\beta +\xi/\epsilon_0)E_{\alpha}(p)}
{\rm e}^{\beta \mu_{\alpha}}
\ee
The factor
\bd
\frac{N!}{N_1! \cdot\cdot\cdot N_n!}
\ed
is just the number of ways to choose $N_1$ particles of type 1, $N_2$ particles of type 2, etc. out of a total of $N=N_1 + \cdot\cdot\cdot +N_n$ particles.  Hence
\be
{\tilde \Zex}(T,\mu,\xi) =  \frac{1}{\xi} \sum_{N=0}^{\infty} \left( \frac{{\tilde z}_1}{\xi} + \cdot\cdot\cdot \frac{{\tilde z}_n}{\xi} \right)^N
= \left( \xi - \sum_{\alpha = 1}^n {\tilde z}_{\alpha} \right)^{-1}
\ee
In the pressure ensemble the pole $\xi_{\rm p}$ furthest to the right along the real axis determines the pressure as $\xi_{\rm p} = \beta \Pex(\beta,\mu)$.  Note that 
\be
\sum_{\alpha = 1}^n {\tilde z}_{\alpha} = \beta_* \Ppt(\beta_*,\mu_*)
\ee
where $\Ppt$ is the point particle pressure with effective inverse temperature $\beta_* = \beta + \xi_{\rm p}/\epsilon_0$ and baryon chemical potential $\mu_* = \beta \mu/\beta_*$.   (The generalization to more than one conserved charge is obvious.)  This implies that the pressure in the excluded volume approximation is expressed in terms of the point particle pressure as
\be
\Pex(T,\mu) = \frac{\Ppt(T_*,\mu_*)}{1-\Ppt(T_*,\mu_*)/\epsilon_0}
\ee
with the real temperature and chemical potential expressed in terms of the effective ones by
\ba
T &=&  \frac{T_*}{1-\Ppt(T_*,\mu_*)/\epsilon_0} \\
\mu &=&  \frac{\mu_*}{1-\Ppt(T_*,\mu_*)/\epsilon_0}
\ea
Straightforward but tedious thermodynamic relations lead to
\ba
\sex(T,\mu) &=& \frac{\spt(T_*,\mu_*)}{1+\ept(T_*,\mu_*)/\epsilon_0} \\
\nex(T,\mu) &=& \frac{\npt(T_*,\mu_*)}{1+\ept(T_*,\mu_*)/\epsilon_0} \\
\eex(T,\mu) &=& -\Pex(T,\mu) + T \sex(T,\mu) + \mu \nex(T,\mu) \nonumber \\
&=& \frac{\ept(T_*,\mu_*)}{1+\ept(T_*,\mu_*)/\epsilon_0}
\ea
Note that in this model there is a natural limiting energy density of $\epsilon_0$.  The model is solved by picking specific values for $T_*$ and $\mu_*$, calculating the point particle properties with these values, and using them to calculate the true $T$ and $\mu$ and thermodynamic properties in the excluded volume approximation.  The chemical potential for each species has the same multiplicative factor.

It is rather tedious to present the derivation with quantum statistics.  The result is simply to calculate the point particle quantities with the inclusion of Bose or Fermi statistics.  The fundamental thermodynamic relations may easily be checked.

It is instructive to take the nonrelativistic limit with one species of particle with mass $m$ and with classical statistics.  Using $\Ppt = \npt T_*$, $\ept = (m + \tthalf T_*)\npt$, and assuming $T \ll m$ and $\nex T \ll \epsilon_0$, one finds the standard van der Waals equation of state $\Pex (1 - v_0 \nex) = \nex T$ where $v_0 = m/\epsilon_0$.

\subsection{Model II}

Now let us consider a different version of the excluded volume approximation where a hadron species $\alpha$ has volume $v_{\alpha}$.  This is referred to as model II.  Following the same procedure as for model I we find
\be
{\tilde z}_{\alpha} = (2s_{\alpha}+1) \int \frac{d^3p}{(2\pi)^3} {\rm e}^{-\beta E_{\alpha}(p)}
{\rm e}^{\beta (\mu_{\alpha}-v_{\alpha} T \xi)}
\ee
Thus the chemical potential for species $\alpha$ is shifted by the amount
\be
\mu_{\alpha} \rightarrow \bar{\mu}_{\alpha} = \mu_{\alpha} - v_{\alpha} T \xi_{\rm p} = \mu_{\alpha} - v_{\alpha} \Pex(T,\mu)
\ee
Thus the pressure must be calculated self-consistently from the equation
\be
\Pex(T,\mu) = \sum_{\alpha = 1}^n P^{\rm pt}_{\alpha}(T, \bar{\mu}_{\alpha})
\ee
where $\Ppt^{\alpha}(T, \bar{\mu}_{\alpha})$ is the point particle pressure for species $\alpha$ with effective chemical potential $\bar{\mu}_{\alpha}$.  Then the application of standard thermodynamic identities yields the following expressions.
\ba
\nex(T,\mu) &=& \frac{\sum_{\alpha} b_{\alpha} n_{\alpha}^{\rm pt}(T,\bar{\mu}_{\alpha})}{1 + \sum_{\alpha} v_{\alpha} n_{\alpha}^{\rm pt}(T,\bar{\mu}_{\alpha}) } \\
\sex(T,\mu) &=& \frac{\sum_{\alpha} s_{\alpha}^{\rm pt}(T,\bar{\mu}_{\alpha})}{1 + \sum_{\alpha} v_{\alpha} n_{\alpha}^{\rm pt}(T,\bar{\mu}_{\alpha}) } \\
\eex(T,\mu) &=& -\Pex(T,\mu) + T \sex(T,\mu) + \mu \nex(T,\mu) \nonumber \\
 &=& \frac{\sum_{\alpha} \epsilon_{\alpha}^{\rm pt}(T,\bar{\mu}_{\alpha})}{1 + \sum_{\alpha} v_{\alpha} n_{\alpha}^{\rm pt}(T,\bar{\mu}_{\alpha}) } 
\ea
One must pay attention to the notation used above: $n_{\alpha}^{\rm pt}(T,\bar{\mu}_{\alpha})$ is the number density of particles of species $\alpha$ treated as non-interacting point particles, whereas the baryon density for point particles is $\sum_{\alpha} b_{\alpha} n_{\alpha}^{\rm pt}(T,{\mu}_{\alpha})$, with $b_{\alpha}$ the baryon number of species $\alpha$.  In this version of the excluded volume model one first calculates the thermodynamic properties with the true $T$ and $\mu$ using the point particle expressions.  Then one solves for the pressure self-consistently and uses it to renormalize the thermodynamic properties.  Note that the chemical potential for each species has the same shift, they are additively modified, not multiplicatively renormalized as in model I.  For example, the effective chemical potential for nucleons is $\mu -  v_N \Pex(T,\mu)$, for anti-nucleons it is $-\mu -  v_N \Pex(T,\mu)$, and for pions of any charge it is $ -  v_{\pi} \Pex(T,\mu)$ (always negative).  Here we choose $v_{\alpha}$ to be proportional to the mass, namely, $v_{\alpha} = m_{\alpha}/\epsilon_0$ where $\epsilon_0$ is a constant.   Then the model should give results very similar to the other excluded volume model where it is proportional to the single particle energy.  

A derivation which includes quantum statistics is straightforward.  The obvious result is that one just calculates the point particle properties with the Bose-Einstein or Fermi-Dirac distributions instead of the Boltzmann distribution. 

\section{Switching from Hadrons to Quarks and Gluons}
\label{sec:switch}

In this section we study the problem of a smooth switching from a purely hadronic equation of state at low temperatures to a purely quark-gluon equation of state at high temperatures.  We will introduce a switching function to accomplish this.  We will also deduce best-fit values for $\epsilon_0$ in the two excluded volume models, which is a physically interesting result in its own right.

The idea of a switching function has been used in atomic and molecular systems for a long time, usually with little success.  For example, it was found in \cite{Woolley}, in the context of the properties of steam, that it is generally impossible to interpolate monotonically all thermodynamic functions over a range where a system has a transition from one phase to another.   While it may be straightforward to make a switching function from a free energy $f_1(T)$ to $f_2(T)$, either its first or second derivative will deviate greatly from any kind 
of weighted average of the derivatives of $f_1$ and $f_2$ alone.  Sometimes this is physical: in a first-order phase transition, the first derivative of the free energy $\partial f/\partial T$ is discontinuous at the transition temperature, corresponding to a discontinuous change in heat capacity as well as a latent heat for the phase transition. However, lattice QCD calculations show no such discontinuities, at least for zero chemical potential, and the switching has to be done with great care.

We begin by constructing a pressure $P$ which includes a hadronic piece $P_h$, a perturbative QCD piece $P_{qg}$, and a switching function $S$.
\be
P(T,\mu) = S(T,\mu) P_{qg}(T,\mu)  + \left[1 - S(T,\mu)  \right] P_h(T,\mu)
\ee
Here $P_h$ may be computed with any of the three hadronic models (pt, exI, exII).  The switching function must approach zero at low temperatures and chemical potentials and approach one at high temperatures and chemical potentials.  The switching function must also be very smooth to avoid introducing first, second, or higher-order phase transitions.  We choose the following functional form
\ba
 S(T, \mu) &=& \exp\{-\theta(T, \, \mu)\} \nonumber \\  
\theta(T, \mu) &=& \left[ \left( \frac{T}{T_0}\right)^r +   \left(\frac{\mu}{\mu_0} \right)^r  \right]^{-1} 
\ea
with integer $r$.   This function is infinitely differentiable, and goes to zero faster than any power of $T$ as $T \rightarrow 0$ (when $\mu = 0$).  It has three parameters.  However, we will choose $\mu_0 = 3 \pi T_0$.  There are two reasons for this choice.  First, it is consistent with Eq. (\ref{scaleM}).  Second, the crossover region at $\mu = 0$ occurs around $T=170$ MeV, whereas the crossover or phase transition is estimated to occur around $\mu = 1.25$ GeV when $T=0$; see \cite{KapustaEOS}.

The other thermodynamic variables must be calculated from the pressure.
\be
 s =  S s_{qg} + \left( 1 - S \right) s_h + \frac{r \theta^2}{T} \left(\frac{T}{T_0}\right)^r \left( P_{qg} - P_h \right)  S
\ee
\be
 n =  S n_{qg} + \left( 1 - S \right) n_h + \frac{r \theta^2}{\mu} \left(\frac{\mu}{\mu_0}\right)^r \left( P_{qg} - P_h \right)  S
\ee
\be
\epsilon = -P + Ts + \mu n
\ee
We now have two parameters in the switching function, two parameters in the perturbative QCD equation of state, and one parameter in the excluded volume equation of state ($\epsilon_0$ not necessarily the same for both models).

\begin{figure}[thp]
 \includegraphics[width=0.8\linewidth]{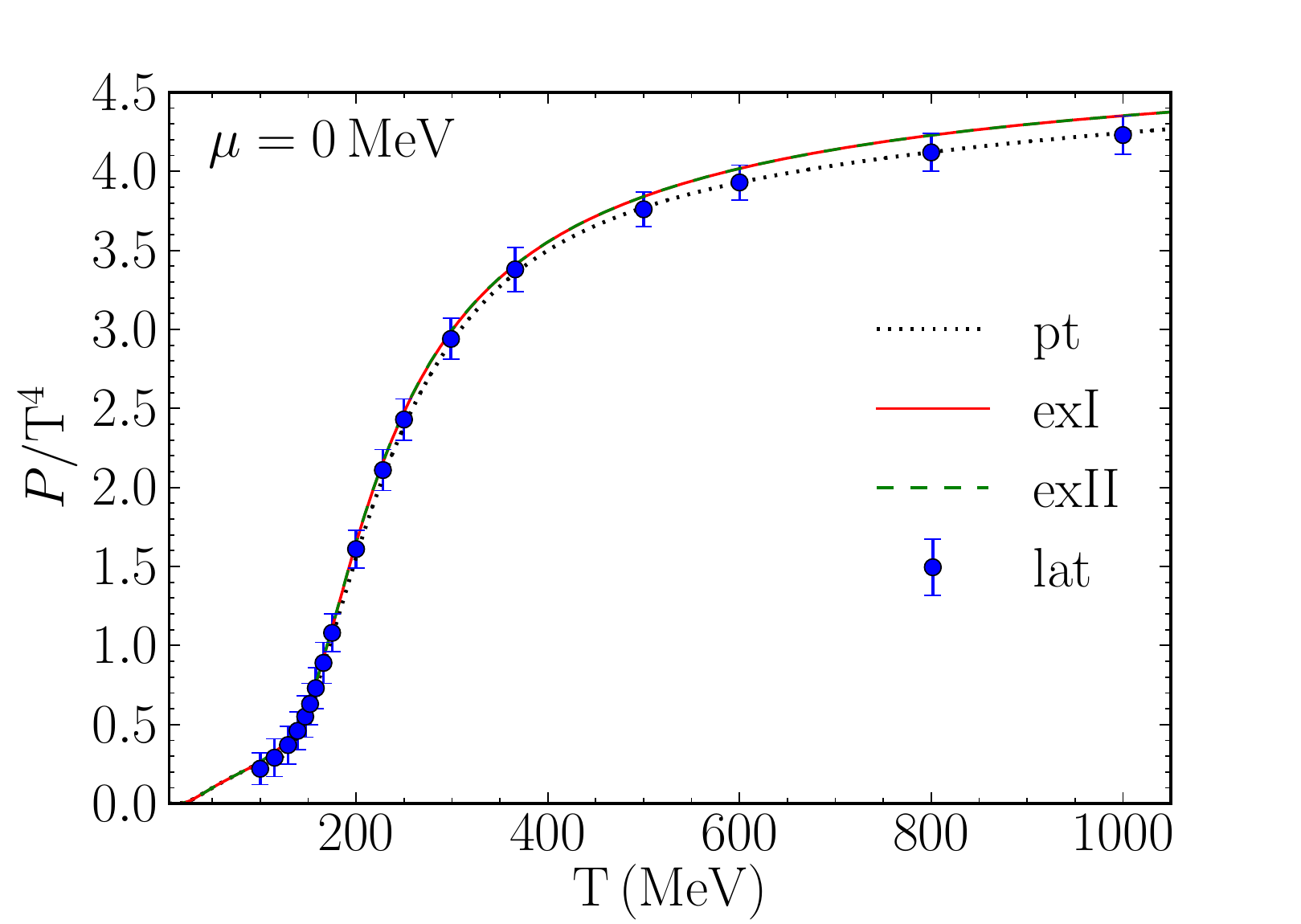} 
 \caption{Pressure of crossover models using best-fit parameters from
 Table~\ref{tb:crossoverfit}.  Lattice data is from \cite{Borsanyi2010}.}
 \label{fig:Pressure0}
\end{figure}

\begin{figure}[thp]
 \includegraphics[width=0.8\linewidth]{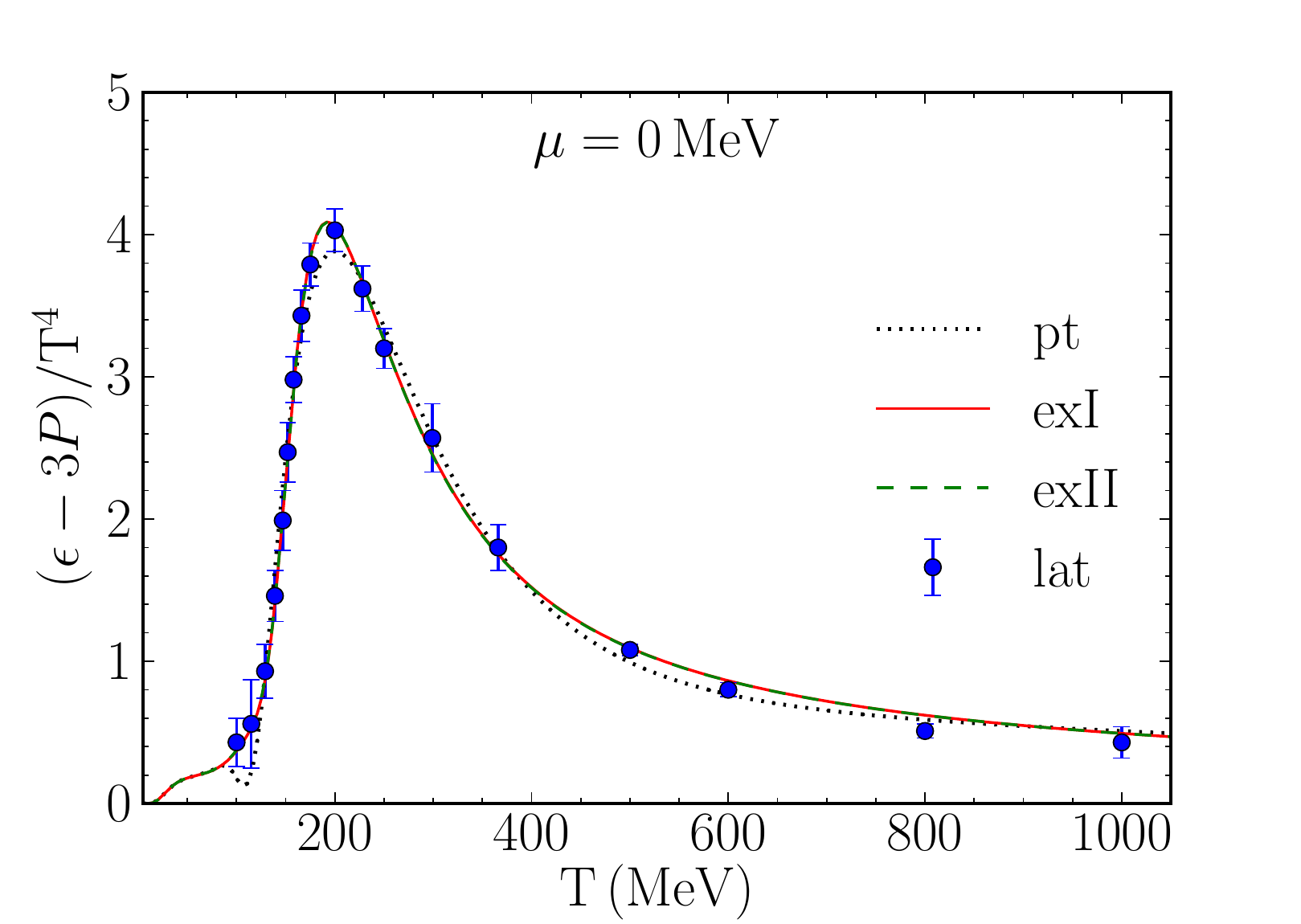} 
 \caption{Trace anomaly of crossover models using best-fit parameters from
 Table~\ref{tb:crossoverfit}.  Lattice data is from \cite{Borsanyi2010}.}
 \label{fig:TraceAnomaly0}
\end{figure}

\begin{figure}
\includegraphics[width=0.8\textwidth]{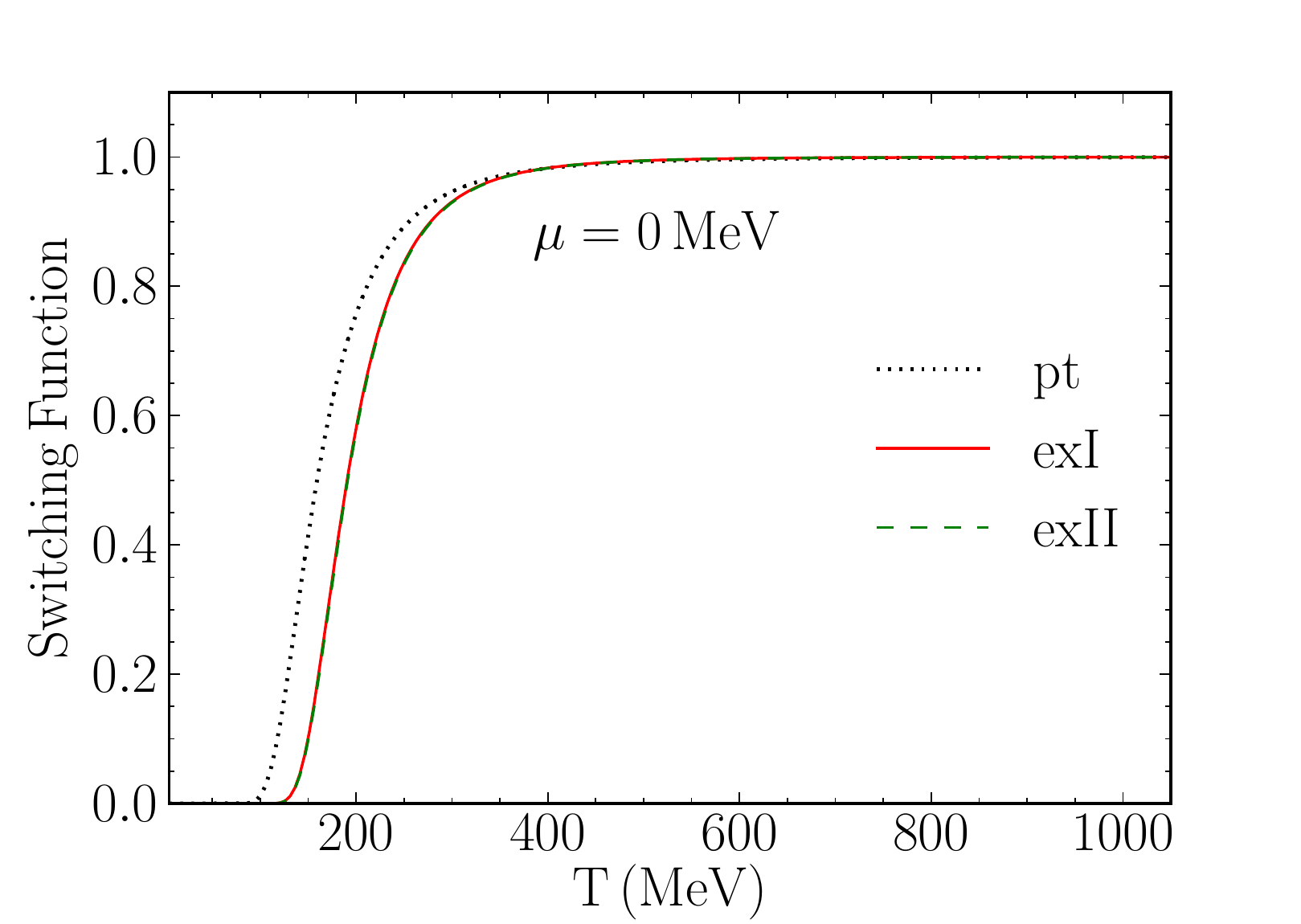}
\caption{Switching function.}
\label{switch}
\end{figure}

We now do a search on the parameters in each of the three models to obtain the best overall chi-square fit to both the pressure and the trace anomaly.  Quantum statistics are used for the hadronic piece of the equation of state.  The results of the fit are shown in Figs. \ref{fig:Pressure0} and \ref{fig:TraceAnomaly0}.  The switching function is shown in Fig. \ref{switch}, and the best fit parameters are shown in Table \ref{tb:crossoverfit}.   

\begin{table}
\begin{tabular}{|c||c|c|c|c|c|c|}
 \hline 
     & $\epsilon_0^{1/4} \, ({\rm MeV})$ & r & $T_0 \, ({\rm MeV})$ & $C_S$  & $C_M$  & $\chi^2 / {\rm dof}$ \\ \hline \hline
 pt   & NA                                    & 4 & 145.33              & 4.196 & 2.855  & 0.558                \\ \hline
 pt   & NA                                    & 5 & 157.44              & 3.896 & 2.965  & 0.616                \\ \hline
 exI  & 306.50                             & 5 & 177.12              & 4.281 & 3.352 & 0.342                \\ \hline
 exI  & 342.27                             & 4 & 175.21              & 1.573  & 3.614 & 0.461                \\ \hline
 exII & 279.71                             & 5 & 177.65              & 4.325 & 3.351 & 0.343                \\ \hline
 exII & 316.28                             & 4 & 175.33             & 1.510  & 3.608 & 0.457 \\ \hline               
 \end{tabular}
 \caption{First and second best-fit parameters for switching function equations
 of state built with pt, exI, and exII hadronic models. Fitting was
 done at $\mu = 0$ with lattice data from \cite{Borsanyi2010}.
 The last column gives each fit's $\chi^2$ per degree of freedom.}
 \label{tb:crossoverfit}
\end{table}

\clearpage

Some points to remark on follow.
\begin{itemize}
\setlength{\itemsep}{0pt}
\item[(a)]  There is essentially no noticeable difference between the model I and model II curves.  The only physical difference between these models is whether the volume excluded by a hadron is proportional to its total energy or to its mass.  Since the hadrons only make a significant contribution below about 200 MeV, the only particle that is really impacted by this difference are the pions, and they contribute only a small part of the total hadronic pressure. 
\item[(b)]  In excluded volume model I, $\epsilon_0$ is the limiting energy density as $T$ becomes large while the pressure increases linearly with $T$.  In model II, both the energy density and the pressure grow slightly faster than $T$.  Hence $P_h/T^4 \sim 1/T^3$ at high temperature.  When multiplied by $1-S$ the hadrons contribute much less than the quarks and gluons, which behave approximately as $P_{qg}/T^4 \sim $ constant.
\item[(c)]  The best fit for model I is obtained with $\epsilon_0 = 1.149$ GeV/fm$^3$ and for model II it is $\epsilon_0 = 797$ MeV/fm$^3$.   These can be used to infer the hard core radius of the proton or neutron to be 0.580 fm for model I and 0.655 fm for model II, very sensible numbers.
\item[(d)]  For the point hadron gas model the best fit is obtained with $r=4$ while the second best fit is obtained with $r=5$.  For the excluded volume models it is just the opposite.  However, the difference in the chi-square between those two values of $r$ is very small. 
\item[(e)]  The value of $T_0$ for the point hadron gas is about 30 MeV smaller than for the excluded volume models.  Thus the switching from hadrons to quarks and gluons occurs at a lower temperature.  The reason is that $P_h$ for the point hadron model grows much faster with $T$ than for the excluded volume models; see Fig. 1 and point (b).  That fast growth must be cut-off by the switching function.  An unnatural consequence is that there is a minor dip in the trace anomaly near a temperature of 115 MeV.
\end{itemize}

\section{Nonzero Chemical Potential}
\label{chemical}

\begin{figure}[h*]
 \includegraphics[width=0.8\textwidth]{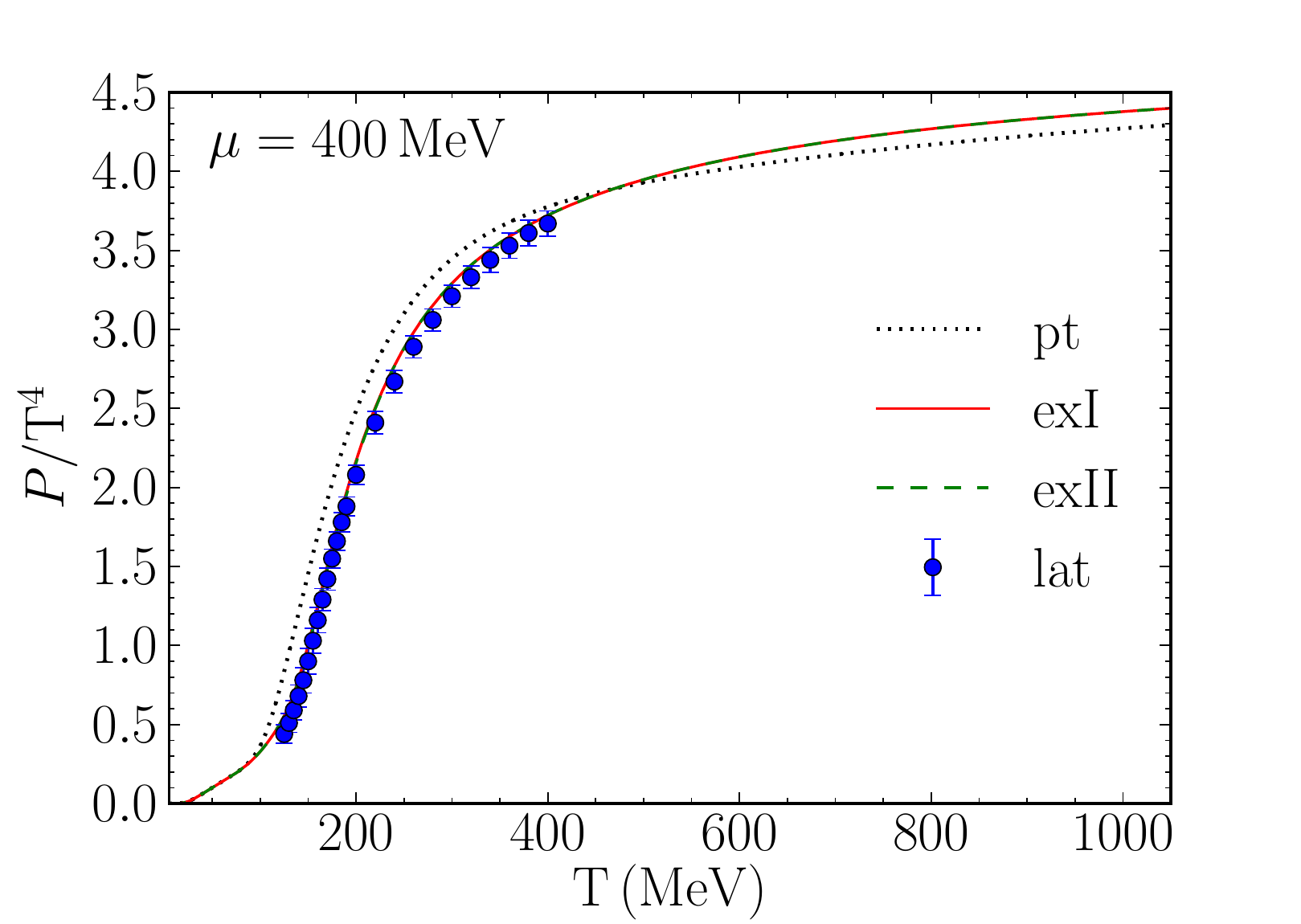} 
 \caption{Pressure of crossover models. 
 Lattice data is from \cite{Borsanyi2012}.}
 \label{fig:Pressure400}
\end{figure}

\begin{figure}
 \includegraphics[width=0.8\textwidth]{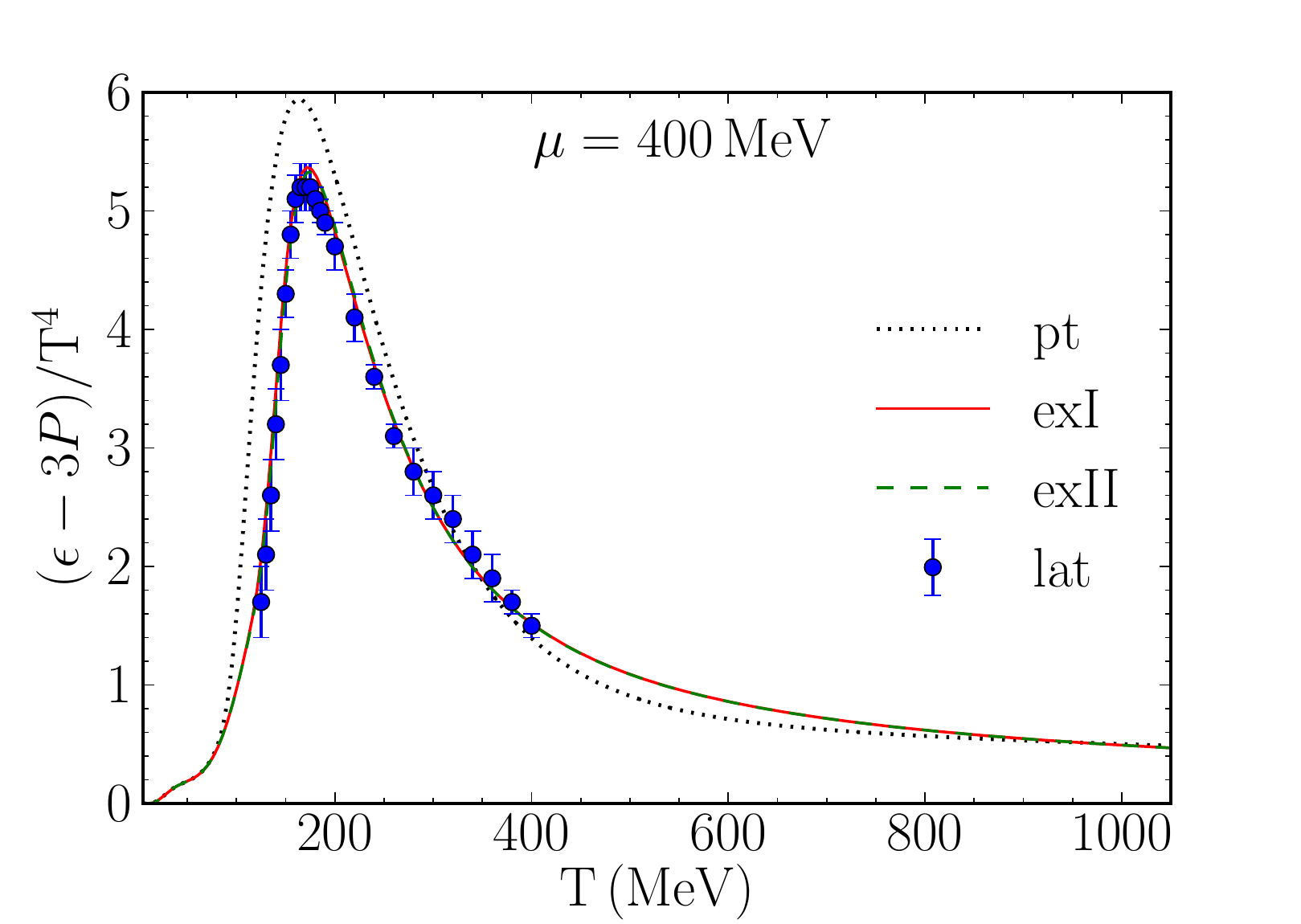} 
 \caption{Trace anomaly of crossover models.  
 Lattice data is from \cite{Borsanyi2012}.}
 \label{fig:TraceAnomaly400}
\end{figure}

The equation of the state can be computed for nonzero baryon chemical potential.  Comparisons are made with lattice results for $\mu = 400$ MeV in Figs. \ref{fig:Pressure400} and \ref{fig:TraceAnomaly400}.  Again the two excluded volume models agree very well with the lattice results.  The model with point hadrons does not agree as well.  It should be emphasized that there are no free parameters in making these comparisons.  All parameters were fixed already.

\begin{figure}
 \includegraphics[width=0.8\textwidth]{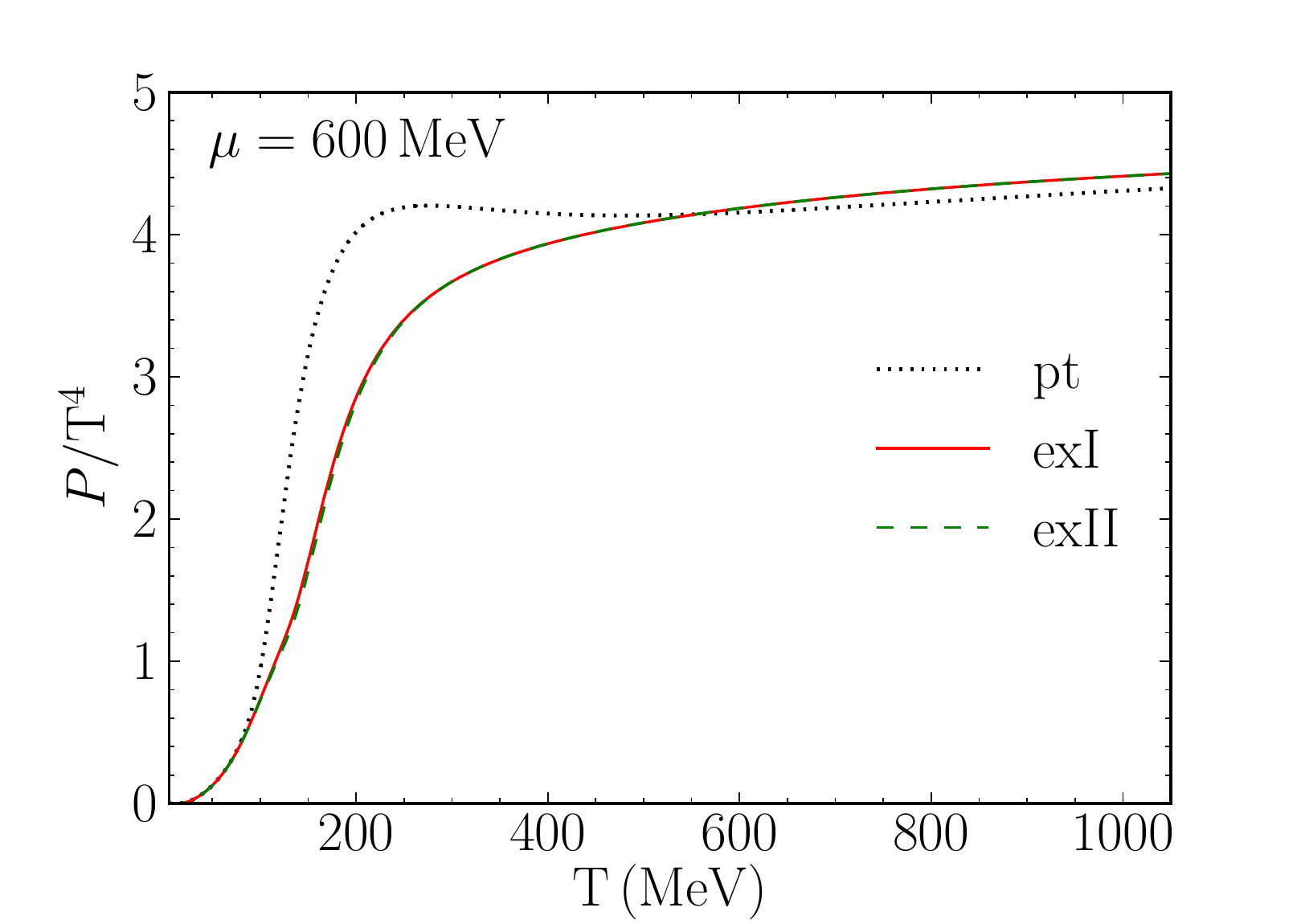} 
 \caption{Pressure of crossover models.}
 \label{fig:Pressure600}
\end{figure}

\begin{figure}
 \includegraphics[width=0.8\textwidth]{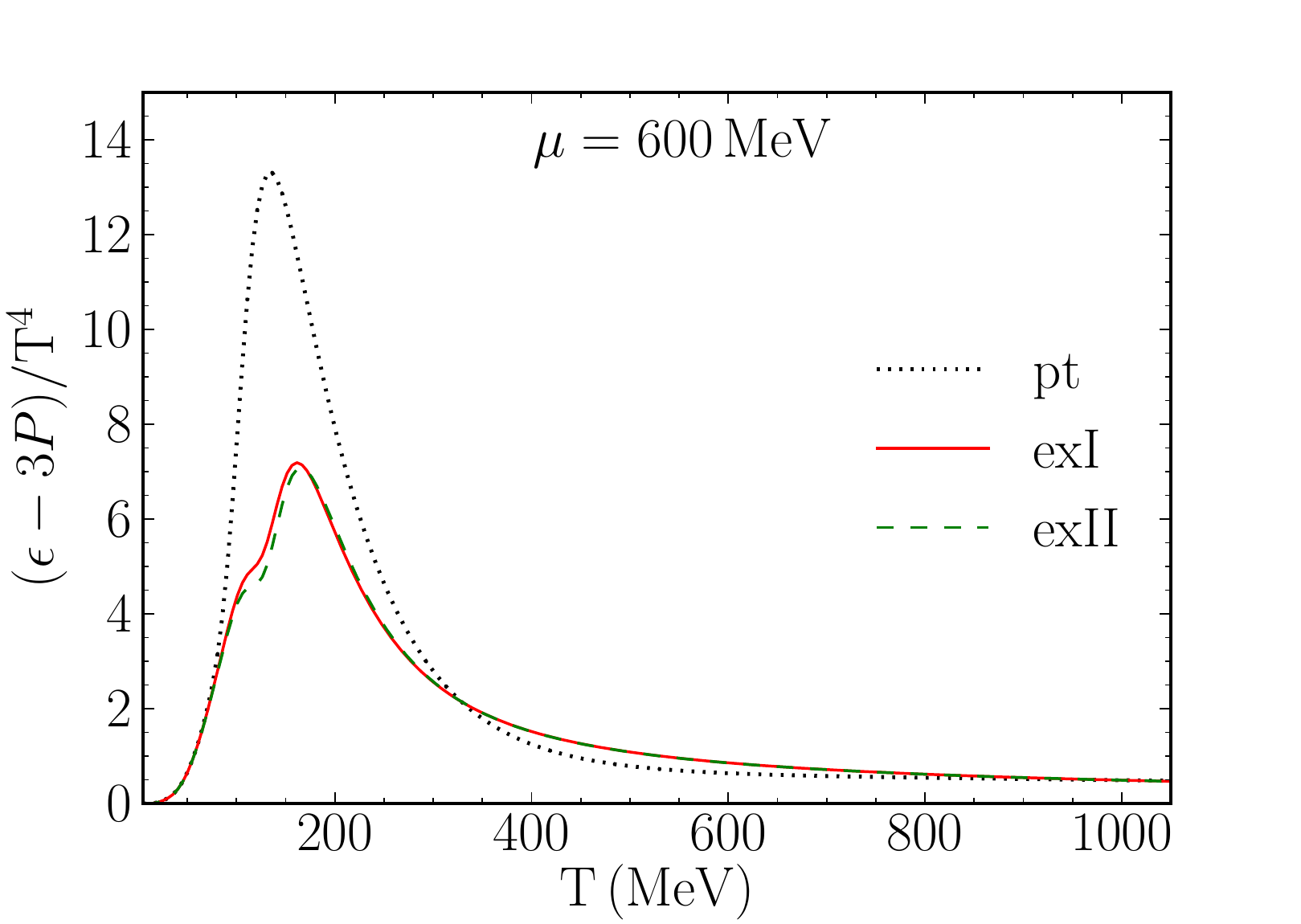} 
 \caption{Trace anomaly of crossover models.}
 \label{fig:TraceAnomaly600}
\end{figure}

In Figs. \ref{fig:Pressure600} and \ref{fig:TraceAnomaly600} we show our results for the larger value $\mu = 600$ MeV.  The difference between the two excluded volume models continues to be insignificant, but now there are large - factor of 2 - differences between them and the point hadron model in the vicinity of $T = 150$ MeV.  



\clearpage

\section{Conclusion}
\label{conclusion}

In this paper we matched three semi-realistic hadronic equations of state at low energy densities to a perturbatively computed equation of state of quarks and gluons at high energy densities.  All three hadronic equations of state include all known hadronic resonances, which approximates attractive interactions among hadrons.  The other two include, in addition, repulsive interactions in the form of excluded volumes occupied by hadrons of finite spatial extent.  A switching function was employed to make the crossover transition from one phase to another without introducing a thermodynamic phase transition.  A chi-square fit to accurate lattice calculations at zero chemical potentials, with temperatures $100 < T < 1000$ MeV, fixes the various parameters in the models.  These parameters quantify the behavior of the QCD running gauge coupling and the physical size of hadrons.  Notably, the hard core radius of protons and neutrons turns out to be $0.62 \pm 0.04$ fm, a very sensible range that lends credence to the models.

The most physically reasonable models include the excluded volume effect.  Not only do they include the effects of attractive and repulsive interactions among hadrons, but they also represent the lattice results the best.  As pointed out by \cite{Woolley}, it is very important to make the best possible approximation to the equation of state in two different phases when attempting to match them, especially when there is no true thermodynamic phase transition, but only a crossover.  

The equations of state constructed in this paper do not result in a phase transition, at least not for the temperatures and baryon chemical potentials investigated.  It remains to be seen how well these equations of state will represent experimental data on high energy heavy ion collisions when implemented in hydrodynamic simulations.

\section*{Acknowledgements}

This work was supported by the US Department of Energy (DOE) under Grant No. DE-FG02-87ER40328.

\newpage

\section*{Appendix A}

This appendix contains a listing of the hadrons included in our calculations.  They are taken from the Particle Data Group.  There are a few baryons whose spins are not known; in these cases we conservatively take them to be spin 1/2.  This table does not include hadrons with charm, bottom, or top quarks, and is therefore the appropriate set of particles for comparisons of equations of state with lattice QCD results including only up, down and strange quarks. The degeneracies of the hadrons includes isospin degeneracy when the mass splitting is small (for example, for the $\Delta$ baryons); otherwise the hadrons are listed separately (for example, $p$ and $n$).  Anti-baryons are not listed.

\begin{table}[h]
%vertical spacing: 1 is normal, < 1 shrinks
\def\arraystretch{.7}
%font: \tiny \scriptsize \footnotesize \small \normalsize
\footnotesize
\begin{tabularx}{\textwidth}{b b s s | b b s s | b b s s} 
\hline \hline
hadron & $m_{\alpha}(\rm GeV)$ & degen & $b_{\alpha}$ & hadron & $m_{\alpha}(\rm GeV)$ & degen & $b_{\alpha}$ & hadron & $m_{\alpha}(\rm GeV)$ & degen & $b_{\alpha}$ \\ 
\hline
 $\pi^0$ & $0.135$ & $1$ & $0$ &  $K^{*0}_2(1430)$ & $1.432$ & $10$ & $0$ &  $K_3^*(1780)$ & $1.776$ & $28$ & $0$ \\ 
 $\pi^{\pm}$ & $0.140$ & $2$ & $0$ &  $N(1440)$ & $1.440$ & $4$ & $1$ &  $\Lambda(1800)$ & $1.800$ & $2$ & $1$ \\ 
 $K^{\pm}$ & $0.494$ & $2$ & $0$ &  $\rho(1450)$ & $1.465$ & $9$ & $0$ &  $\Lambda(1810)$ & $1.810$ & $2$ & $1$ \\ 
 $K^0$ & $0.498$ & $2$ & $0$ &  $a_0(1450)$ & $1.474$ & $3$ & $0$ &  $\pi(1800)$ & $1.812$ & $3$ & $0$ \\ 
 $\eta$ & $0.548$ & $1$ & $0$ &  $\eta(1475)$ & $1.476$ & $1$ & $0$ &  $K_2(1820)$ & $1.816$ & $20$ & $0$ \\ 
 $\rho$ & $0.775$ & $9$ & $0$ &  $f_0(1500)$ & $1.505$ & $1$ & $0$ &  $\Lambda(1820)$ & $1.820$ & $6$ & $1$ \\ 
 $\omega$ & $0.783$ & $3$ & $0$ &  $\Lambda(1520)$ & $1.520$ & $4$ & $1$ &  $\Xi(1820)$ & $1.823$ & $8$ & $1$ \\ 
 $K^{*\pm}(892)$ & $0.892$ & $6$ & $0$ &  $N(1520)$ & $1.520$ & $8$ & $1$ &  $\Lambda(1830)$ & $1.830$ & $6$ & $1$ \\ 
 $K^{*0}(892)$ & $0.896$ & $6$ & $0$ &  $f_2^{'}(1525)$ & $1.525$ & $5$ & $0$ &  $\phi_3(1850)$ & $1.854$ & $7$ & $0$ \\ 
 $p$ & $0.938$ & $2$ & $1$ &  $\Xi^0(1530)$ & $1.532$ & $4$ & $1$ &  $N(1875)$ & $1.875$ & $8$ & $1$ \\ 
 $n$ & $0.940$ & $2$ & $1$ &  $N(1535)$ & $1.535$ & $4$ & $1$ &  $\Delta(1905)$ & $1.880$ & $24$ & $1$ \\ 
 $\eta^{'}$ & $0.958$ & $1$ & $0$ &  $\Xi^-(1530)$ & $1.535$ & $4$ & $1$ &  $\Delta(1910)$ & $1.890$ & $8$ & $1$ \\ 
 $a_0$ & $0.980$ & $3$ & $0$ &  $\Delta(1600)$ & $1.600$ & $16$ & $1$ &  $\Lambda(1890)$ & $1.890$ & $4$ & $1$ \\ 
 $f_0$ & $0.990$ & $1$ & $0$ &  $\Lambda(1600)$ & $1.600$ & $2$ & $1$ &  $\pi_2(1880)$ & $1.895$ & $15$ & $0$ \\ 
 $\phi$ & $1.019$ & $3$ & $0$ &  $\eta_2(1645)$ & $1.617$ & $5$ & $0$ &  $N(1900)$ & $1.900$ & $8$ & $1$ \\ 
 $\Lambda$ & $1.116$ & $2$ & $1$ &  $\Delta(1620)$ & $1.630$ & $8$ & $1$ &  $\Sigma(1915)$ & $1.915$ & $18$ & $1$ \\ 
 $h_1$ & $1.170$ & $3$ & $0$ &  $N(1650)$ & $1.655$ & $4$ & $1$ &  $\Delta(1920)$ & $1.920$ & $16$ & $1$ \\ 
 $\Sigma^+$ & $1.189$ & $2$ & $1$ &  $\Sigma(1660)$ & $1.660$ & $6$ & $1$ &  $\Delta(1950)$ & $1.930$ & $32$ & $1$ \\ 
 $\Sigma^0$ & $1.193$ & $2$ & $1$ &  $\pi_1(1600)$ & $1.662$ & $9$ & $0$ &  $\Sigma(1940)$ & $1.940$ & $12$ & $1$ \\ 
 $\Sigma^-$ & $1.197$ & $2$ & $1$ &  $\omega_3(1670)$ & $1.667$ & $7$ & $0$ &  $f_2(1950)$ & $1.944$ & $5$ & $0$ \\ 
 $b_1$ & $1.230$ & $9$ & $0$ &  $\omega(1650)$ & $1.670$ & $3$ & $0$ &  $\Delta(1930)$ & $1.950$ & $24$ & $1$ \\ 
 $a_1$ & $1.230$ & $9$ & $0$ &  $\Lambda(1670)$ & $1.670$ & $2$ & $1$ &  $\Xi(1950)$ & $1.950$ & $4$ & $1$ \\ 
 $\Delta$ & $1.232$ & $16$ & $1$ &  $\Sigma(1670)$ & $1.670$ & $12$ & $1$ &  $a_4(2040)$ & $1.996$ & $27$ & $0$ \\ 
 $K_1(1270)$ & $1.272$ & $12$ & $0$ &  $\pi_2(1670)$ & $1.672$ & $15$ & $0$ &  $f_2(2010)$ & $2.011$ & $5$ & $0$ \\ 
 $f_2$ & $1.275$ & $5$ & $0$ &  $\Omega^-$ & $1.673$ & $4$ & $1$ &  $f_4(2050)$ & $2.018$ & $9$ & $0$ \\ 
 $f_1$ & $1.282$ & $3$ & $0$ &  $N(1675)$ & $1.675$ & $12$ & $1$ &  $\Xi(2030)$ & $2.025$ & $12$ & $1$ \\ 
 $\eta(1295)$ & $1.294$ & $1$ & $0$ &  $\phi(1680)$ & $1.680$ & $3$ & $0$ &  $\Sigma(2030)$ & $2.030$ & $24$ & $1$ \\ 
 $\pi(1300)$ & $1.300$ & $3$ & $0$ &  $N(1680)$ & $1.685$ & $12$ & $1$ &  $K_4^*(2045)$ & $2.045$ & $36$ & $0$ \\ 
 $\Xi^0$ & $1.315$ & $2$ & $1$ &  $\rho_3(1690)$ & $1.689$ & $21$ & $0$ &  $\Lambda(2100)$ & $2.100$ & $8$ & $1$ \\ 
 $a_2$ & $1.318$ & $15$ & $0$ &  $\Lambda(1690)$ & $1.690$ & $4$ & $1$ &  $\Lambda(2110)$ & $2.110$ & $6$ & $1$ \\ 
 $\Xi^-$ & $1.322$ & $2$ & $1$ &  $\Xi(1690)$ & $1.690$ & $4$ & $1$ &  $\phi(2170)$ & $2.175$ & $3$ & $0$ \\ 
 $f_0(1370)$ & $1.350$ & $1$ & $0$ &  $N(1700)$ & $1.700$ & $8$ & $1$ &  $N(2190)$ & $2.190$ & $16$ & $1$ \\ 
 $\pi_1(1400)$ & $1.354$ & $9$ & $0$ &  $\Delta(1700)$ & $1.700$ & $16$ & $1$ &  $N(2200)$ & $2.250$ & $20$ & $1$ \\ 
 $\Sigma(1385)$ & $1.385$ & $12$ & $1$ &  $N(1710)$ & $1.710$ & $4$ & $1$ &  $\Sigma(2250)$ & $2.250$ & $6$ & $1$ \\ 
 $K_1(1400)$ & $1.403$ & $12$ & $0$ &  $K^*(1680)$ & $1.717$ & $12$ & $0$ &  $\Omega^-(2250)$ & $2.252$ & $2$ & $1$ \\ 
 $\Lambda(1405)$ & $1.405$ & $2$ & $1$ &  $\rho(1700)$ & $1.720$ & $9$ & $0$ &  $N(2250)$ & $2.275$ & $20$ & $1$ \\ 
 $\eta(1405)$ & $1.409$ & $1$ & $0$ &  $f_0(1710)$ & $1.720$ & $1$ & $0$ &  $f_2(2300)$ & $2.297$ & $5$ & $0$ \\ 
 $K^*(1410)$ & $1.414$ & $12$ & $0$ &  $N(1720)$ & $1.720$ & $8$ & $1$ &  $f_2(2340)$ & $2.339$ & $5$ & $0$ \\ 
 $\omega(1420)$ & $1.425$ & $3$ & $0$ &  $\Sigma(1750)$ & $1.750$ & $6$ & $1$ &  $\Lambda(2350)$ & $2.350$ & $10$ & $1$ \\ 
 $K^*_0(1430)$ & $1.425$ & $4$ & $0$ &  $K_2(1770)$ & $1.773$ & $20$ & $0$ &  $\Delta(2420)$ & $2.420$ & $48$ & $1$ \\ 
 $K^{*\pm}_2(1430)$ & $1.426$ & $10$ & $0$ &  $\Sigma(1775)$ & $1.775$ & $18$ & $1$ &  $N(2600)$ & $2.600$ & $24$ & $1$ \\ 
 $f_1(1420)$ & $1.426$ & $3$ & $0$ &  & & &  &  & & &  \\ 
\hline \hline 
\end{tabularx}
\end{table}

\newpage

\section*{Appendix B}

In this appendix we summarize the perturbative QCD equation of state used to describe the quark-gluon plasma phase.  These are obtained from refs. \cite{Vuorinen2003} and \cite{Haque2013}.   Note that both $f_4$ and $f_6$ depend on $\ln(\alpha_s/\pi)$.

\ba
 P =  \frac{8 \pi^2}{45} T^4 \left[ f_0  + f_2  \left({\alpha_s \over \pi}\right)
 + f_3  \left( {\alpha_s\over \pi} \right)^{3/2} + f_4  \left( {\alpha_s \over \pi} \right)^2
 + f_5  \left( {\alpha_s \over \pi} \right)^{5/2}  \!\! + f_6\left( {\alpha_s \over \pi} \right)^3 \right]
 \label{eq:PQCD:Pressure}
\ea 
where 
\ba
f_0 &=& 1 + \frac{3 N_f}{32}\left(7+120\hat\mu_q^2 +240\hat\mu_q^4 \right) 
 \\
f_2 &=& - {15 \over 4} \left[ 1 + {N_f\over 12}\left(5+72\hat\mu_q^2 +144\hat\mu_q^4 \right)  \right]
 \\
f_3 &=& 30 \left[ 1 + \frac{N_f}{6}\left(1 + 12\hat\mu_q^2\right) \right]^{3/2} \\
f_4 &=&  237.223 + \left(15.963 + 124.773\ \hat\mu_q^2 -319.849\hat\mu_q^4 \right) N_f \nonumber \\
&-&\left( 0.415 + 15.926\ \hat\mu_q^2 + 106.719\ \hat\mu_q^4\right)N_f^2 \nonumber  \\
&+& { 135 \over 2} \left[ 1 + \frac{N_f}{6}\left(1+12\hat\mu_q^2\right) \right]
\ln \left[ \left({\alpha_s \over \pi}\right) \left(1 + \frac{N_f}{6}\left(1+12\hat\mu_q^2\right) \right) \right] \nonumber \\
&-&{165\over 8} \left[1+\frac{N_f}{12}\left(5+72\hat\mu_q^2 +144\hat\mu_q^4 \right) \right]\left(1 -\frac{2N_f}{33} \right)\ln{\hat M} \\ \nonumber
 f_5 &=& -\sqrt{ 1 + \frac{N_f}{6}\left(1+12\hat\mu_q^2\right)}
 	\Bigg[ 799.149 + \left(21.963 - 136.33\ \hat \mu_q^2 + 482.171\ \hat\mu_q^4 \right)N_f 
 \nonumber \\
 &+& \left(1.926 + 2.0749\ \hat\mu_q^2 - 172.07\ \hat\mu_q^4\right) N_f^2\Bigg] 
 \nonumber\\
&+&\ {495\over 12} \left(6+ N_f (1+12\hat \mu_q^2) \right)\left(1 -\frac{2N_f}{33}\right)\ln{\hat M} \\
 f_6 &=& -\Bigg[659.175 + \left(65.888 -341.489\ \hat\mu_q^2 + 1446.514\ \hat\mu_q^4\right)N_f 
  \nonumber \\
 &+& \left(7.653 + 16.225\ \hat \mu_q^2 - 516.210\ \hat \mu_q^4\right) N_f^2
 \nonumber\\
 &-& \frac{1485}{2}\left(1+\frac{1+12\hat\mu_q^2}{6}N_f\right)\left(1-\frac{2N_f}{33} \right)\ln{\hat M} \Bigg]
 \ln \left[ \left(\frac{\alpha_s}{\pi}\right) \left(1+\frac{N_f}{6}(1+12 \hat \mu_q^2) \right)4\pi^2\right]\nonumber \\
 &-&  475.587\ln \left[ \left(\frac{\alpha_s}{\pi}\right) 4\pi^2C_A \right] 
\ea
For QCD we have $N_c = 3$, $C_A = 3$, and we take $N_f = 3$.  The $M$ is the renormalization scale.  If $\mu$ is the baryon chemical potential then $\mu_q = \mu/3$.  The hat denotes division by $2 \pi T$ so that $\hat\mu_q = \mu_q/(2 \pi T)$ and $\hat M = M/ (2 \pi T)$.

\newpage

We use the 3-loop coupling constant from the PDG \cite{pdg2012} (we drop the $b_3$ term).
\ba
 \label{eq:alpha_s}
 \alpha_s &=& \frac{1}{b_0 t} \left[ 
   1 - \frac{b_1}{b_0^2}\frac{\ln t}{t}
  + \frac{b_1^2(\ln^2 t - \ln t - 1) + b_0b_2 }{b_0^4 t^2} \right. \nonumber \\
  && \left. - \frac{b_1^3(\ln^3 t - \frac52 \ln^2 t - 2\ln t + \frac12)
      + 3b_0b_1b_2 \ln t  }{b_0^6 t^3} \right] 
\ea
where
\bd
 b_0 = \frac{33-2N_f}{12\pi}
\ed
\bd
 b_1 = \frac{153-19N_f}{24\pi^2}
\ed
\be
 b_2 = \frac{1}{128 \pi^3} \left(2857 - {5033 \over 9} N_f + {325 \over 27} N_f^2 \right)
\ee
We make one modification to $t$: we introduce a constant $C_S$ to soften its divergence at low temperatures and chemical potentials.
\be
 t \equiv \ln \left(C_S^2 + M^2/\Lambda^2_{\overline{MS}} \right)
\ee
We recover the equation of state of \cite{Haque2013} when $C_S = 0$, and take $\Lambda_{\overline{MS}} = 290$ MeV as in that paper.

\end{document}